# A generalized Hausdorff distance based quality metric for point cloud geometry

*Alireza Javaheri, Catarina Brites, Fernando Pereira and João Ascenso*

Instituto Superior Técnico – Universidade de Lisboa, Instituto de Telecomunicações, Lisbon, Portugal

{alireza.javaheri, catarina.brites, fp, joao.ascenso}@lx.it.pt

*Abstract*— Reliable quality assessment of decoded point cloud geometry is essential to evaluate the compression performance of emerging point cloud coding solutions and guarantee some target quality of experience. This paper proposes a novel point cloud geometry quality assessment metric based on a generalization of the Hausdorff distance. To achieve this goal, the so-called generalized Hausdorff distance for multiple rankings is exploited to identify the best performing quality metric in terms of correlation with the MOS scores obtained from a subjective test campaign. The experimental results show that the quality metric derived from the classical Hausdorff distance leads to low objective-subjective correlation and, thus, fails to accurately evaluate the quality of decoded point clouds for emerging codecs. However, the quality metric derived from the generalized Hausdorff distance with an appropriately selected ranking, outperforms the MPEG adopted geometry quality metrics when decoded point clouds with different types of coding distortions are considered.

## I. INTRODUCTION

Point cloud (PC) quality assessment plays an important role in many emerging applications and also on the effective design and development of novel PC coding solutions. In immersive communications, very large PCs are common and, thus, it is critical to compress them efficiently before transmission and storage. In this context, objective quality assessment metrics are used to evaluate the decoded PCs quality for a specific target rate, thus determining the rate-distortion (RD) performance for the coding solution. Currently, the so-called D1 and D2 geometry PSNR quality metrics are adopted by MPEG [1] and are largely used as state-of-the-art quality metrics for PC geometry quality assessment. Although the PSNR computed from the classical Hausdorff distance shows a high objective-subjective correlation performance for octree-based PC codecs [2][3], MPEG no longer uses this metric for PC geometry quality assessment due to its low reliability for the emerging MPEG PC coding standards, notably Geometry-based Point Cloud Compression (G-PCC) [4] and Video-based Point Cloud Compression (V-PCC) [5], which are based on other coding paradigms. In fact, the classical Hausdorff distance based geometry PSNR quality metric is very sensitive to outlier points in the decoded PC, which may not even be visible (or rendered), as the relevant metric distance corresponds to the greatest of all the distances from a point in one PC to the closest point in the other PC (original to decoded, and vice-versa). This extreme outlier sensitive behavior was the motivation to perform a statistical analysis of the distances' distribution for PCs coded with different codecs, to find a more reliable quality metric for PC geometry. In this context, this paper proposes a more reliable PC geometry quality metric, notably in terms of objective-subjective correlation, which is based on the so-called generalized Hausdorff (GH) distance [6]. The generalized Hausdorff distance is an extension of the classical Hausdorff distance that measures the distance between two PCs using the $K^{th}$ ranked distance instead of the last ranked (i.e. maximum) distance; this is equivalent to the last ranked (maximum) distance in a (K-values) portion of the sorted distances. While it has been previously used for object matching in noisy images [7], the generalized Hausdorff distance has never been used for PC quality assessment. Since a small number of points with large errors/distances (i.e. outlier points) dispersed in a decoded PC may not be even very visible, the generalized Hausdorff distance may be a better choice for PC geometry quality assessment since outlier points may be discarded by considering only the $K^{th}$ lowest distances. In this paper, the generalized Hausdorff distance parameter (i.e. the rank K) is varied to obtain 15 different distances from which quality metrics are obtained with some pooling function; after, their objective-subjective correlation performances are used to identify the distance (and pooling function) leading to the best associated PSNR quality assessment metric, notably for PCs coded with the state-of-the-art MPEG PC standard codecs, i.e. G-PCC and V-PCC, and a simple codec based on pure octree pruning available in the Point Cloud Library (PCL) codec [8]. In summary, this paper proposes a novel PC geometry quality metric outperforming the state-of-the-art as represented by the adopted MPEG PC geometry quality metrics.

The rest of this paper is organized as follows. Section II presents the state-of-the-art objective quality metrics related with the goal of this paper, while Section III describes the proposed PC geometry quality assessment metric. The experimental results are presented and analyzed in Section IV. Finally, some conclusions are described in Section V.

## II. STATE-OF-THE-ART POINT CLOUD GEOMETRY QUALITY ASSESSMENT

In this section, the related work is reviewed, namely relevant PC geometry quality metrics in the literature able to deal with various compression artifacts as well as the adopted MPEG PC geometry quality metrics, which will be taken as reference.

### A. Beyond MPEG PC Geometry Quality Metrics

The PC quality assessment field has seen recently many contributions both in terms of subjective evaluation methodologies and objective quality metrics. To evaluate the performance and reliability of objective metrics to assess decoded PC quality, subjective assessment experiments need to be performed first. Many subjective studies available in the literature rely on artifacts generated from Gaussian noise or octree pruning, which is a very simple coding scheme with a rather distinctive type of distortions [9]-[12]. In [13], a large study on the impact of rendering on the perceived quality of decoded PCs using the emerging MPEG PC coding standards is presented. In [14], a subjective assessment of MPEG codecs (as well as the MPEG adopted objective metrics) considering decoded geometry and color is performed while [15] describes a subjective test on dynamic PC data coded with V-PCC.

Regarding PC objective quality metrics, they can be classified in two main categories: point-based metrics and projection-based metrics. While point-based metrics consider the correspondence between points in the original and decoded PC, projection-based metrics map the 3D PCs onto more classical 2D planes. The most popular point-based geometry objective metrics for PC quality assessment are the point-to-point metrics [16], which depend on the distance between corresponding points, the point-to-plane metrics [17], which depend on the distance between a point and the tangent plane at the corresponding point, and the plane-to-plane geometry metrics [18], which depend on the similarity between tangent planes at corresponding points. PSNR metrics based on point-to-point and point-to-plane distances are currently the state-of-the-art metrics and are used in the MPEG standardization activities (described in in the next section). Furthermore, in [19], a structural similarity-based PC geometry quality metric is proposed based on the local curvature statistics. This metric computes first curvatures for each point and establishes after correspondences between points. The distortion measure corresponds to Gaussian weighted curvature statistics on a set of local neighborhoods. In [20], a projection-based metric is proposed by applying an orthographic projection to obtain 2D color maps. Then, a 2D image quality metric such as PSNR, SSIM [21] or VQM [22] is used to assess the quality. This metric considers both geometry and color simultaneously.

*B. MPEG PC Geometry Quality Metrics*

Since the MPEG adopted PC geometry quality metrics represent currently the state-of-the-art, they will be here presented in more detail. Considering the distance between two points $a$ and $b$ as $d(a,b) = \|a - b\|_2^2$, the distance between the point $a$ in PC $A$ and the PC $B$ is defined as:

$$d_{P2Po}(a, B) = \|\vec{e}_{a,B}\|_2^2 = \min_{b \in B} d(a, b) \quad (1)$$

corresponding to the directed point-to-point distance from $a$ to the nearest neighboring point in PC $B$; in (1), $\vec{e}_{a,B}$ stands for the error vector resulting from connecting point $a$ to the nearest neighbor point in PC $B$. With the point-to-point distance definition, several *point-to-point* (*P2Po*) quality metrics may be defined with appropriate pooling and normalization, e.g. as currently done for the PSNR metric which will be defined later. The distance between a point and a reference plane, where this plane is a coarse representation of the surface around a point, is also very often used, leading to the so-called *point-to-plane* (*P2Pl*) quality metrics. In this case, the point-to-point error $\vec{e}_{a,B}$ (1) is projected along the normal direction to the underlying surface at a given point in PC $B$, thus becoming:

$$d_{P2Pl}(a, B) = (\vec{e}_{a,B} \cdot \vec{n}_b)^2 \quad (2)$$

where $\vec{n}_b$ is the normal vector at the $a$ nearest neighbor point in PC $B$. The directed (i.e. for one specific direction) distance between two PCs can be measured in different ways in the two directions, this means from the original to the decoded PCs and vice-versa. The MPEG adopted PC geometry quality metrics use *MSE* for the two directed distances, notably:

$$d_{MSE}(A, B) = \frac{1}{N_A}\sum_{a \in A} d(a, B) \quad (3)$$

$$d_{MSE}(B, A) = \frac{1}{N_B}\sum_{b \in B} d(b, A) \quad (4)$$

where $N_A$ and $N_B$ are the total number of points in PC A and PC B, respectively. Both point-to-point and point-to-plane distances can be used to compute $d_{MSE}$ in (3) and (4). Previously, MPEG was also using the classical Hausdorff distance as directed distance defined as:

$$d_{Haus}(A, B) = \max_{a \in A} d(a, B). \quad (5)$$

This distance is not used anymore due to its poor objective-subjective correlation performance for PCs coded with the emerging MPEG PC standard codecs, notably V-PCC.

The undirected (i.e. symmetric) distance between two PCs $A$ and $B$ can be measured by pooling the directed distances from both directions. Different pooling functions can be used, but the maximum function defined as in (6) is rather common to measure the undirected distance.

$$pool(d(A, B), d(B, A)) = \max(d(A, B), d(B, A)) \quad (6)$$

Finally, a PSNR may be obtained from the undirected distance, e.g. MSE based distance $d_{MSE}$, between PCs as:

$$PSNR = 10 \log_{10}(\frac{3p^2}{d_{MSE}}) \quad (7)$$

where $p$ is the signal peak which normalizes the error and $d_{MSE}$ is the undirected distance, which can be obtained by using (3) and (4) in (6). Equivalently, the undirected PSNR can also be computed by pooling with the minimum function, *min*(), over the directed PSNRs obtained from directed MSE distances. In MPEG, the point-to-point MSE based PSNR metric is called D1 PSNR while the point-to-plane MSE based PSNR metric is called D2 PSNR [1].

III. PROPOSED PC GEOMETRY QUALITY METRIC

The classical Hausdorff distance is very sensitive to outliers, since one or more points with a large error magnitude will dominate the final quality score even for cases where these points are not much visible. The classical Hausdorff distance defined in Section II can be generalized to compute the distance over a subset of data/distances after ranking all the distance/error values. Instead of taking the maximum distance over all the distances as in the classical Hausdorff distance, the generalized Hausdorff distance for rank K is computed using only the K lowest distance values after ranking all the distances in ascending order. Thus, the K[th] ranked generalized Hausdorff distance is defined as:

$$d_{GH-K}(A, B) = {}^{per}K^{th}_{a \in A} d(a, B) \quad (8)$$

where ${}^{per}K^{th}_{a \in A}$ is the K[th] ranked distance such that $(K/N_A) \times 100 = per\%$ and $N_A$ is the total number of points in PC $A$. For example, the 480[th] ranked distance in a PC with 600 points is the maximum distance obtained from the $per$=(480/600)×100=80% lowest distance values, after sorting all the distances in ascending order.

In this paper, 15 directed generalized Hausdorff distances are obtained by assigning different values to *per%* in the generalized Hausdorff distance as defined in (8), where $d_{100/N_a}$ and $d_{100}$ are special cases corresponding to:

$$d_{\frac{100}{N_a}}(A,B) = \min_{a \in A} d(a,B) = {}^{\frac{100}{N_A}}K_{a \in A}^{th} d(a,B) \quad (9)$$

$$d_{100}(A,B) = \max_{a \in A} d(a,B) = {}^{100}K_{a \in A}^{th} d(a,B) \quad (10)$$

Note that $d_{100}$ corresponds to the classical Hausdorff distance mentioned before. Moreover, four different pooling functions are considered to compute the undirected distance (i.e. considering both directions) between two PCs, notably:

$$pool_{min}(d(A,B), d(B,A)) = \min(d(A,B), d(B,A)) \quad (11)$$

$$pool_{max}(d(A,B), d(B,A)) = \max(d(A,B), d(B,A)) \quad (12)$$

$$pool_{avg}(d(A,B), d(B,A)) = \frac{d(A,B) + d(B,A)}{2} \quad (13)$$

$$pool_{wavg}(d(A,B), d(B,A)) = \frac{N_A d(A,B) + N_B d(B,A)}{N_A + N_B} \quad (14)$$

By combining the 15 directed distances with the four proposed undirected distance pooling functions, 60 undirected distances are obtained, labelled in the following as $D_{per,pool}$, which will be used to measure the distortion between two PCs (original and decoded). After, the corresponding PSNR quality metric is computed for all these undirected distances, using them as a substitute to $d_{MSE}$ in (7) while using the same $p$. For example, $PSNR_{65,wavg}$ is obtained from the undirected distance $D_{65,wavg}$, which results from applying the weighted average pooling function to pool the directed distances $d_{65}$ from both directions associated to $per = 65\%$; note that $PSNR_{100,max}$ corresponds to the classical Hausdorff distance based PSNR. In the following, all these PSNR quality metrics will be assessed to identify the best performing ones.

## IV. PERFORMANCE EVALUATION

This section assesses the performance of the proposed quality metrics for different ranks and pooling functions to find the best performing PC geometry quality metric for PCs decoded by different PC codecs, notably the MPEG standard codecs.

### A. Subjective Test Characterization

In this paper, the MOS scores and PCs available in the Rendered Point Cloud (IRPC) quality assessment dataset [23] are used since they correspond to very extensive and representative conditions. The original and decoded PCs are compared and evaluated by subjects using a Double Stimulus Impairment Scale (DSIS) subjective assessment protocol. The adopted PC dataset includes six PCs from the MPEG repository [1], notably *Egyptian Mask* and *Frog* from Inanimate Objects; *Facade9* and *House without roof*, from Façades and Buildings; and *Loot* and *Longdress* from People. The key characteristics of these PCs are listed in Table I.

The selected PCs have been coded with three rates/qualities using the following PC codecs: i) octree-based codec from PCL [8]; ii) MPEG G-PCC [4]; and iii) MPEG V-PCC codec [5]. The selected codecs represent three different coding paradigms, notably PCL for tree structures, MPEG G-PCC for surface models (when *trisoup* is used) and MPEG V-PCC for projection-based coding. The bitrates for each codec were selected from the rates defined in the MPEG Common Test Conditions (CTC) [1], to have three distinguishable qualities for each PC. The coding parameters used for the PCL, MPEG G-PCC and V-PCC codecs are the same as in [13].

TABLE I. TEST MATERIAL AND RESPECTIVE CHARACTERISTICS.

| PC Name | No. Points | Precision | Category | Signal Peak ($p$) |
|---|---|---|---|---|
| *Egyptian Mask* | 272,684 | 12 bit | Inanimate Objects | 4095 |
| *Frog* | 3,614,251 | 12 bit | Inanimate Objects | 4095 |
| *Facade9* | 1,596,085 | 12 bit | Facades & Buildings | 4095 |
| *House without roof* | 4,848,745 | 12 bit | Facades & Buildings | 4095 |
| *Loot* | 805,285 | 10 bit | People | 1023 |
| *Longdress* | 857,966 | 10 bit | People | 1023 |

The decoded PCs have been subjectively assessed using a DSIS subjective assessment protocol in three sessions, each corresponding to a different rendering approach, notably *RPoint* (rendering using a point representation with uniform color and shading), *RMesh* (rendering using a mesh representation with uniform color and shading), and *RColor* (rendering using a point representation and original colors). Because there are no coloring and interpolation processes involved in *RPoint* rendering, geometry coding artifacts are less masked in this rendering approach [13]. Since the proposed PC quality metrics described in Section III only evaluate PC geometry, only the *RPoint* session MOS scores [23] will be used in the following. There were 20 subjects participating in each assessment session and no outliers were found for the *RPoint* session, whose MOS scores are used in the experiments presented next.

### B. Experimental Results

As usual, the performance of a quality metric is measured by computing the correlation of the metric scores with respect to human opinion scores. For PC geometry quality assessment, this implies that, even when all points are not exactly in the same positions in the two PCs *A* and *B* under comparison, as long as *A* and *B* look perceptually similar, a good metric should create a score close to zero. Thus, the proposed ranked distances should discard distances/errors that are not visible, i.e. perceptually relevant, to approximate this thresholding effect of the human visual system. Moreover, to obtain reliable objective quality assessment metrics, the adopted distances need to discriminate well between different perceptual quality levels for PC geometry. Several experiments with decoded PCs using the three previously selected PC codecs have shown that $D_{100/N,-}$ and $D_{50,-}$ are 0 most of the time, especially for high qualities. Moreover, $D_{60,min}$, $D_{65,min}$ and $D_{70,min}$ are also often 0 for V-PCC coding, clearly indicating that they do not have enough discriminatory power to evaluate V-PCC decoded PCs. However, only $D_{100/N,-}$ to $D_{50,-}$ are not further used in the following performance assessment exercise, which targets identifying the best performing (PSNR) quality metric and the corresponding generalized Hausdorff distance rank.

To evaluate the objective-subjective correlation performance for the proposed PC geometry quality metrics and associated distances, non-linear regression has been used to map the computed objective quality scores (PSNR) into the MOS scores available from the DSIS subjective test [23]. In this case, the following cubic function was used:

$$MOS_p = a + by + cy^2 + dy^3 \quad (15)$$

where $y$ are the objective metric scores and $a$, ..., $d$ are regression model parameters [24].

The Pearson Linear Correlation Coefficient (PLCC), as a measure of the linear dependence between objective metric scores and subjective (opinion) scores, the Spearman Ranked Order Correlation Coefficient (SROCC), as a measure of the strength and direction of monotonicity between the two scores abovementioned, and the Root Mean Square Error (RMSE), as a measure of the dispersion of the objective scores with respect to the subjective scores, have been selected to assess the objective-subjective correlation.

Fig. 1 shows the overall PLCC performance for the point-to-point and point-to-plane PSNR computed for the generalized Hausdorff distance using different ranks and pooling functions in comparison with the *MPEG D1* and *MPEG D2* (PSNR) metrics by considering the MOS scores from all selected codecs together. The results show that above $per = 90\%$ ($d_{90}$), the generalized Hausdorff based PSNR PLCC starts to outperform the *MPEG D1* and *D2* PSNR PLCC when the three selected PC codecs are considered together. There is also a significant drop in PLCC from the generalized Hausdorff based PSNR to the classical Hausdorff based PSNR (extreme right side of the charts), thus highlighting that a small number of dispersed large distances/errors (outliers) are not visible; in fact, the PLCC charts show that their consideration on the objective metrics largely penalizes the objective-subjective correlation. The best PLCC performing point-to-point metric is $PSNR_{98,avg}$, corresponding to $D_{98,avg}$, which considers 98% of the data/distances; for point-to-plane metrics, the best performing metric is $PSNR_{99,min}$, corresponding to $D_{99,min}$ where 99% of data/distances are used. Although the experimental results do not show a significant PLCC difference between the various pooling functions, for point-to-point distances, the $pool_{min}$ function is clearly the best for lower $per\%$ while the $pool_{avg}$ function is the best for high $per\%$. For point-to-plane distances, the opposite behavior can be observed.

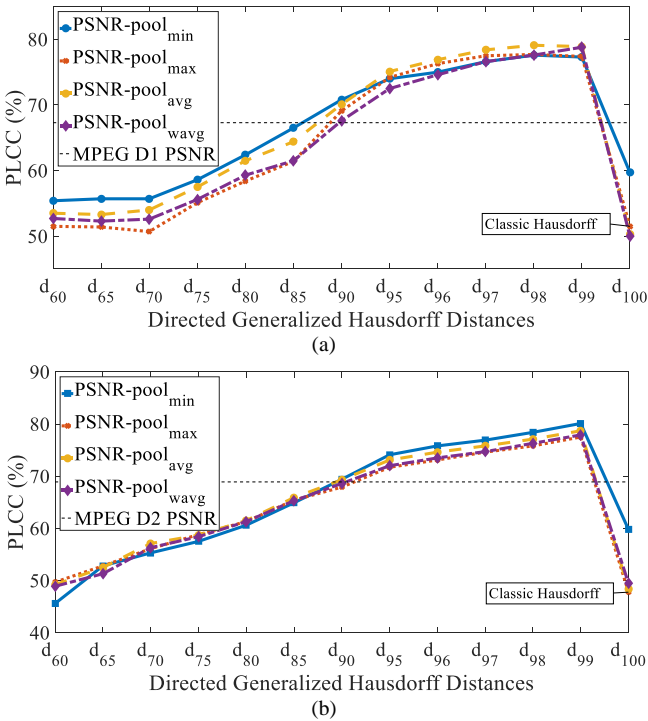

Fig. 1. PLCC performance for the (a) point-to-point and (b) point-to-plane generalized Hausdorff PSNR for different poolings as a function of the directed rank distance in comparison with MPEG D1 and D2 (PSNR) quality metrics.

The PLCC results in Fig. 1 for $per = 100\%$ ($d_{100}$) clearly indicates the presence of outliers in the distances' distribution, which strongly penalize the correlation. Fig. 2 shows the ranked distance for different percentage of data/distances ($per$ in (8)), individually for the PCL, G-PCC and V-PCC codecs. The top row in Fig. 2 shows the distances from original to decoded PCs while the bottom row shows the distances from decoded to original PCs. For G-PCC and V-PCC, there is a large and sudden increase of the maximum distance, close to $per = 100\%$ while, for PCL, the distances increase much slower. This behavior highlights that for G-PCC and V-PCC the decoded PCs have a very small portion of distances/errors much larger than the average distance/error. This behavior also explains why the classical Hausdorff distance performs well for PCL [2][3] but not for G-PCC and V-PCC, thus justifying the MPEG to not adopt this metric anymore.

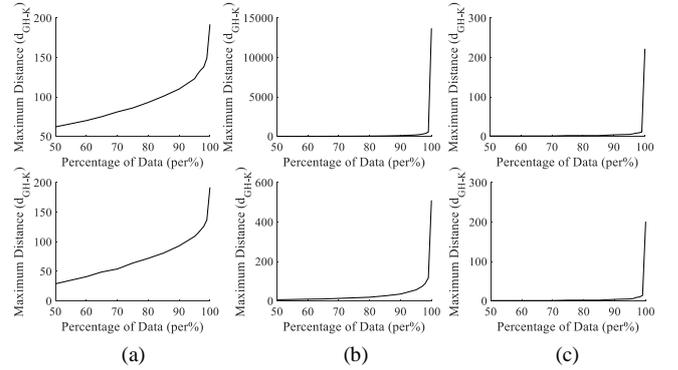

Fig. 2. Maximum of directed squared distances from original to decoded on top and decoded to original on bottom plotted for each selected percentage of data for three codecs: (a) PCL; (b) G-PCC; (c) V-PCC.

To assess the proposed quality metrics improvements regarding the MPEG metrics for each of the selected PC codecs, Fig. 3 shows the PLCC between MOS scores and point-to-point generalized Hausdorff PSNR with max pooling, before applying the fitting function in (15). For PCL, the generalized Hausdorff PSNR metric performance with different percentage of data, is similar to MPEG D1 PSNR, which considers all distances/errors. This means that the errors introduced by PCL coding do not include a significant number of outliers. However, for the other two codecs, by not considering some of the distances (i.e. outliers), large correlation performance improvements can be achieved, notably for $d_{95}$ to $d_{98}$.

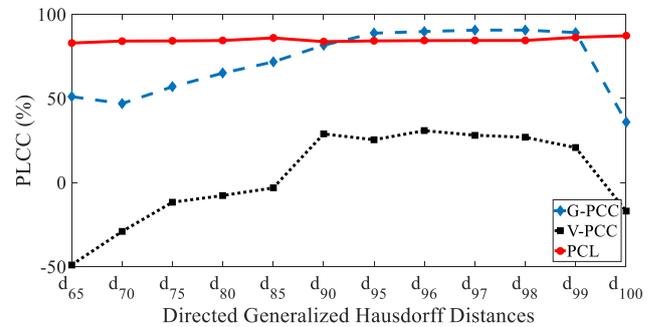

Fig. 3. PLCC performance for point-to-point generalized Hausdorff PSNR using maximum function pooling and D1 PSNR for the three selected PC codecs. Note the D1 - PCL and D1 - G-PCC curves are on top of each other.

Fig. 4 shows the distribution of the coding errors over the *Egyptian Mask* PC coded with PCL, G-PCC and V-PCC. As it can be seen from Fig. 4, the G-PCC and V-PCC decoded PCs have a very small amount of errors whose magnitude is much larger than the average error magnitude (being even invisible at naked eye) and those errors are dispersed over the decoded PCs. Regarding the PCL decoded PC, most of the error magnitudes are around the average error magnitude, which, once more, indicates that the number of outliers (i.e. large errors) introduced by PCL coding is small.

Finally, the PLCC and SROCC performances for the best performing generalized Hausdorff (GH) based PSNR quality metrics, considering each codec individually and altogether, are shown in Table II. These results allow concluding:

- **PCL** – For PCL, the point-to-point generalized Hausdorff based PSNR with rank 100% (classical Hausdorff) outperforms MPEG D1 PSNR by 2.5%, 6.1% and 0.1 for PLCC, SROCC and RMSE, respectively; a similar behavior happens for the point-to-plane generalized Hausdorff with rank 85% ($PSNR_{85,max}$) with gains of 0.3%, 0.9% and 0.1 for PLCC, SROCC and RMSE, respectively, over MPEG D2 PSNR.

- **G-PCC** – For G-PCC, the best performing quality metric is the point-to-point generalized Hausdorff with rank 98% based PSNR ($PSNR_{98,max}$) with gains of 4.4%, 6.1% and 0.1 for PLCC, SROCC and RMSE, respectively, over MPEG D1 PSNR. For point-to-plane, the generalized Hausdorff based PSNR slightly outperforms MPEG D2 PSNR except for SROCC, where a small loss occurs.

- **V-PCC** – For V-PCC, there are major correlation gains compared to the MPEG quality metrics. The best quality metric for V-PCC is the point-to-plane generalized Hausdorff with rank 96% based PSNR ($PSNR_{96,min}$) with gains of 25.2%, 30.3% and 0.2 for PLCC, SROCC and RMSE, respectively, over MPEG D2 PSNR. Large gains also occur for the point-to-point generalized Hausdorff with rank 90% based-PSNR ($PSNR_{98,min}$) over MPEG D1 PSNR. This allows to conclude that the filtering of outliers is more important for this PC codec, especially since a large amount of points is added in the PC decoding.

- **All codecs** – As already shown in Fig. 1, the generalized Hausdorff based PSNR with ranks from 90% to 99% outperform the MPEG quality metrics when decoded data from all codecs are considered together. More precisely, Table II shows that, considering all codecs, the best point-to-point metric is the generalized Hausdorff with rank 98% based PSNR ($PSNR_{98,avg}$) with gains of 11.8%, 13.2% and 0.1 for PLCC, SROCC and RMSE, respectively, over MPEG D1 PSNR; the best point-to-plane metric is the generalized Hausdorff with rank 99% based PSNR ($PSNR_{99,min}$) with gains of 11.2%, 12.6% and 0.1 for PLCC, SROCC and RMSE, respectively, over MPEG D2 PSNR.

As shown in Table II, the generalized Hausdorff distance based PSNR outperforms the MPEG D1 and D2 PSNR metrics, if the right rank (i.e. *per%*) is used. In general, for point-to-point distances, except for data coded with PCL, generalized Hausdorff with 98% of data ($d_{98}$) has the best performance. For point-to-plane metrics, the best generalized Hausdorff varies for each codec. Although not shown in Table II, the PSNR associated with classical Hausdorff distance ($PSNR_{100,max}$) shows poor correlation with MOS scores, except for PCL.

Since it is not the best solution to use different quality metrics for different PC codecs, unless variations of a specific codec are being compared, the correlation performance results have been analyzed to identify the generalized Hausdorff distance based PSNR quality metric which is globally reliable. Although not the optimum for all PC codecs, the point-to-point generalized Hausdorff $PSNR_{96,max}$ shows better correlation compared to the MPEG D1 PSNR metric as shown in Table III for all codecs individually and also altogether. However, there are other generalized Hausdorff based PSNRs that, while not always outperforming the MPEG metric, show very large correlation gains for the codec typically exhibiting lower objective-subjective correlations when using the MPEG metrics, this means V-PCC; however, a small correlation loss may be observed for one of the remaining codecs. As shown in Table III, this is the case for point-to-plane $PSNR_{96,min}$ and point-to-point $PSNR_{98,min}$, which show rather large correlation gains (up to 25.2% and 16.4%, respectively) for V-PCC, the codec where correlation improvements are most needed. In summary, the point-to-plane $PSNR_{96,min}$ is a good metric to outperform the MPEG metrics for the emerging PC coding solutions.

## V. CONCLUSIONS

This paper shows that the generalized Hausdorff based PSNR quality metric outperforms the *MPEG D1* and *D2* PC geometry quality metrics for all considered codecs individually, and also altogether, if the appropriate rank (96%-99%) is selected. In this case, better objective-subjective correlation than the MPEG quality metrics is achieved for point-to-plane $PSNR_{96,min}$, thus indicating that these metrics may be used with advantage for future PC geometry quality assessment.

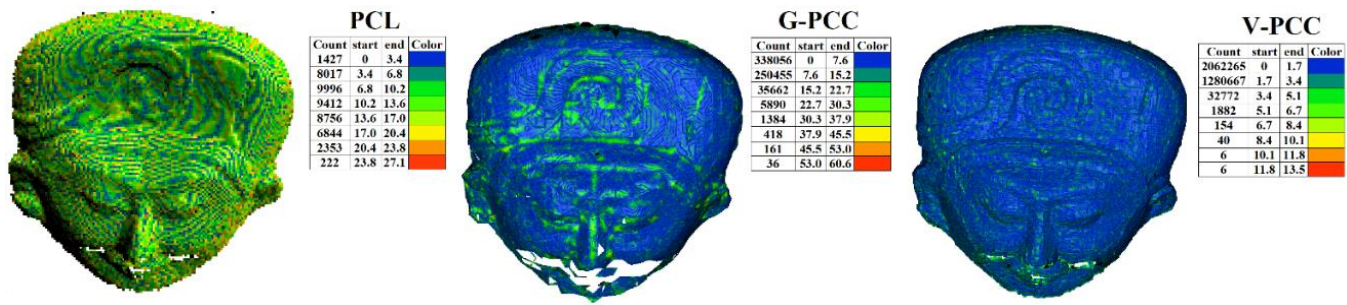

Fig. 4. Distribution of coding errors throughout the decoded *Egyptian Mask* PC for PCL, G-PCC and V-PCC codecs. The tables show the number of occurrences of errors within a certain range, to which a specific color is assigned.

TABLE II. PLCC, SROCC AND RMSE PERFORMANCE FOR THE BEST QUALITY METRIC FOR EACH PC CODEC INDIVIDUALLY AND ALL CODECS TOGETHER.

| | | Point-to-point metrics | | | | Point-to-plane metrics | | | |
|---|---|---|---|---|---|---|---|---|---|
| | | *Quality metric* | *PLCC* | *SROCC* | *RMSE* | *Quality metric* | *PLCC* | *SROCC* | *RMSE* |
| PCL | *Best GH* | $PSNR_{100,max}$ | **89.5** | **79.9** | **0.5** | $PSNR_{85,max}$ | **90.5** | **86.7** | **0.5** |
| | *MPEG* | D1 PSNR | 87.0 | 73.8 | 0.6 | D2 PSNR | 90.2 | 85.8 | 0.6 |
| | *Gain* | - | **2.5** | **6.1** | **0.1** | - | **0.3** | **0.9** | **0.1** |
| G-PCC | *Best GH* | $PSNR_{98,max}$ | **91.3** | **93.5** | **0.4** | $PSNR_{97,min}$ | **89.4** | 91.7 | **0.4** |
| | *MPEG* | D1 PSNR | 86.9 | 87.4 | 0.5 | D2 PSNR | 89.1 | **93.0** | 0.5 |
| | *Gain* | - | **4.4** | **6.1** | **0.1** | - | **0.3** | **-1.3** | **0.1** |
| V-PCC | *Best GH* | $PSNR_{98,min}$ | **69.5** | **67.6** | **0.5** | $PSNR_{96,min}$ | **76.6** | **79.9** | **0.4** |
| | *MPEG* | D1 PSNR | 53.1 | 62.0 | 0.6 | D2 PSNR | 51.4 | 49.6 | 0.6 |
| | *Gain* | - | **16.4** | **5.6** | **0.1** | - | **25.2** | **30.3** | **0.2** |
| All Codecs | *Best GH* | $PSNR_{98,avg}$ | **79.1** | **77.9** | **0.6** | $PSNR_{99,min}$ | **80.1** | **77.7** | **0.6** |
| | *MPEG* | D1 PSNR | 67.3 | 64.7 | 0.7 | D2 PSNR | 68.9 | 65.1 | 0.7 |
| | *Gain* | - | **11.8** | **13.2** | **0.1** | - | **11.2** | **12.6** | **0.1** |

TABLE III. PLCC PERFORMANCE FOR THE BEST OVERALL GENERALIZED HAUSDORFF QUALITY METRICS.

| | Point-to-point metrics | | | | | Point-to-plane metrics | | |
|---|---|---|---|---|---|---|---|---|
| | **MPEG D1 PSNR** | $PSNR_{96,max}$ | **Gain** | $PSNR_{98,min}$ | **Gain** | **MPEG D2 PSNR** | $PSNR_{96,min}$ | **Gain** |
| **PCL** | 87.0 | 87.1 | **0.1** | 89.0 | **2.0** | 90.2 | 89.0 | -1.2 |
| **G-PCC** | 86.9 | 90.3 | **3.4** | 86.5 | -0.4 | 89.1 | 89.2 | **0.1** |
| **V-PCC** | 53.1 | 53.5 | **0.4** | 69.5 | **16.4** | 51.4 | 76.6 | **25.2** |
| **All codecs** | 67.3 | 76.3 | **9.0** | 77.6 | **10.3** | 68.9 | 75.8 | **6.9** |
| **Average** | 73.5 | 76.8 | **3.2** | 80.7 | **7.1** | 74.9 | 82.7 | **7.8** |

ACKNOWLEDGEMENT

This work is funded by FCT/MCTES through national funds and when applicable co-funded EU funds under the project UIDB/EEA/50008/2020 and PTDC/EEI-PRO/7237/2014.